\documentclass[numberedappendix,floatfix, twocolumn]{aastex631}

\usepackage{bm}
\usepackage{xspace}
\usepackage{enumitem}
\usepackage{graphicx}
\usepackage{color}
\usepackage{comment}
\usepackage{amsmath}
\usepackage[T1]{fontenc}
\usepackage{natbib}
\usepackage[capitalize]{cleveref}
\usepackage{colortbl}
\usepackage[toc]{appendix}
\usepackage{soul}
\usepackage{wrapfig}

\Crefmultiformat{equation}{Eqs.~#2#1#3}%
{,~#2#1#3}{, #2#1#3}{ and~#2#1#3}

\Crefformat{equation}{Eq.~#2#1#3}

\usepackage{mathtools}

\newcommand{\orcid}[1]{\href{https://orcid.org/#1}{\includegraphics[width=8pt]{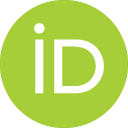}}}

\shorttitle{How to Escape From a Trap}
\shortauthors{Gilbaum et al.}

\begin{document}

\title{How to Escape from a Trap: Outcomes of Repeated Black Hole Mergers in AGN}
\correspondingauthor{Shmuel Gilbaum}
\email{shmuel.gilbaum@mail.huji.ac.il}
\author[0000-0002-6462-6657]{Shmuel Gilbaum}
\affiliation{Racah Institute of Physics, The Hebrew University, Jerusalem, 91904, Israel}
\author[0000-0001-7113-723X]{Evgeni Grishin}
\affiliation{School of Physics and Astronomy, Monash University, Clayton, VIC 3800, Australia}
\affiliation{The ARC Centre of Excellence for Gravitational Wave Discovery – OzGrav, Clayton, VIC 3800, Australia}

\author[0000-0002-4337-9458]{Nicholas C.~Stone}
\affiliation{Racah Institute of Physics, The Hebrew University, Jerusalem, 91904, Israel}
\affiliation{Department of Astronomy, University of Wisconsin, Madison, WI, 53706}

\author[0000-0002-6134-8946]{Ilya Mandel}
\affiliation{School of Physics and Astronomy, Monash University, Clayton, VIC 3800, Australia}
\affiliation{The ARC Centre of Excellence for Gravitational Wave Discovery – OzGrav, Clayton, VIC 3800, Australia}

\begin{abstract}

Stellar-mass black holes (BHs) embedded in active galactic nuclei (AGN) may be major sources of astrophysical gravitational waves (GWs), contributing both to the observed LIGO-Virgo-KAGRA population of binary BH mergers and to future populations of {\it LISA}-band extreme mass ratio inspirals (EMRIs).  The ability of these BHs to pair up into binaries, inspiral, and produce GWs will be shaped by the existence of migration traps, regions in the AGN where hydrodynamic torques vanish.  Previous works have studied the existence and location of migration traps in AGN disks.  Here, we investigate how individual BHs may escape such traps as an outcome of mergers, potentially suppressing hierarchical growth.  We find that while GW recoil kicks are strong enough to kick merged BHs onto inclined orbits, gas drag quickly realigns them into the AGN disk.  A more robust escape mechanism is gap opening: once a BH grows above a critical mass, its gravity disturbs the AGN gas sufficiently to eliminate the trap.  In low-mass AGN relevant for {\it LISA}, gaps open easily and the resulting ``wet EMRI'' masses are unlikely to reflect protracted hierarchical mergers.  In combination with our previous work, we find that migration traps only exist in a relatively narrow range of AGN luminosities, $L \in [10^{43.5},10^{45.5}]~{\rm erg~s}^{-1}$. We identify an even narrower AGN luminosity range for which stellar mass BHs can grow into the pair instability mass gap and beyond. This characteristic luminosity scale may assist in indirect tests of the ``AGN channel'' for binary BH mergers.

\end{abstract}


\section{Introduction}
\label{introduction}

The new era of gravitational wave (GW) astronomy has produced multiple breakthrough detections, from a multimessenger double neutron star merger \citep{Abbott+17a, Abbott+17b} to the recent, tentatively observed stochastic background \citep{Agazie+23, EPTA23, Reardon+23, Xu+23} from supermassive black holes (SMBHs).  However, the overwhelming majority of GW signals detected to date come from stellar mass black holes, which inspiral and merge in binary black hole (BBH) pairs.  Many scenarios have been proposed to explain the BBH populations discovered by the LIGO-Virgo-KAGRA (LVK) collaboration \citep{Abbott+23a, Abbott+23b}, including binary common envelope evolution \citep{TutukovYungelson93, Belczynski+02}; binary chemically homogeneous evolution \citep{MandeldeMink16,Marchant+2016}; chaotic three-body scatterings in dense star clusters \citep{PortegiesZwartMcMillan00, Rodriguez+16}; secular Kozai-Lidov cycles in hierarchical triples \citep{AntoniniPerets12, Antonini+16, Gri18, hoang18, fra19}; and lastly, hydrodynamic effects in the gas-rich environments of active galactic nuclei \citep{McKernan+14, Bellovary+16, Stone+17, Bartos+17a, Tagawa+20}, or AGN \footnote{ See, e.g.,\cite{Mapelli2021,MandelFarmer2022} for reviews of these and other formation channels and \cite{MandelBroekgaarden2022} for a review of the merger rate predictions. }. 

The ``AGN channel,'' which is the focus of this paper, has many uniquely testable features.  For example, the gas-rich environment makes it theoretically possible for BBHs in AGN to produce electromagnetic counterparts \citep{McKernan+19, Tagawa+24}, although outshining the central AGN is a major challenge for observability\footnote{Secondarily, galactic nuclei host a variety of electromagnetic transients, some of which may be confused for a hypothetical BBH merger counterpart \citep{Zabludoff+21, gri+21}}.  The relative scarcity of AGN makes it plausible to statistically test the association of merging BBH with the AGN channel \citep{Bartos+17b, Veronesi+22, Veronesi+23, veronesi24}, something that cannot be done for other BBH formation pathways.  The tremendous escape velocities of AGN permit retention of BBH merger products \citep{Samsing+22}, and the dissipative gas environment allows such BHs to pair up again and grow hierarchically through repeated mergers \citep{Tagawa+20, Tagawa+21,Vaccaro+2024}.  Such repeated mergers allow later generations of BHs to populate the pair instability mass gap \citep{GerosaFishbach21} in the BH mass spectrum (see e.g. GW190521; \citealt{Abbott+20}), thought to be forbidden for single-star evolution \citep{Barkat+67, Takahashi18, Farmer+19}.  

Unfortunately, the AGN channel also features an unusually high level of theoretical uncertainty, largely due to the rich mix of dynamics, hydrodynamics, and stellar evolution involved.  BBHs may form (i) {\it in situ} \citep{Stone+17}, (ii) from the capture of a pre-existing population on inclined orbits \citep{Bartos+17a, GenerozovPerets2023}, or (iii) dynamically, from single-single encounters of individual BHs within the AGN disk \citep{Bellovary+16, Tagawa+20}.  This latter possibility is particularly intriguing, as dynamical binary formation is necessary for hierarchical growth; aside from producing unique LVK masses, hierarchical growth may also leave an imprint on the population of extreme mass ratio inspirals, or EMRIs \citep{HilsBender95, Ryan95}, observable with future space-based GW interferometers like {\it LISA} \citep{AmaroSeoane+23} and {\it Taiji/TianQin} \citep{Luo+16, Ruan+20}.  Binary assembly in AGN disks may in principle occur at all radii, driven by a combination of gas dynamical friction \citep{Tagawa+20} and circum-single disk interactions \citep{Li+23, Rowan+23, Whitehead+24} dissipating low relative velocities in close flybys, but it will be especially efficient in {\it migration traps}.

Individual BHs or other stellar mass objects embedded in AGN disks will migrate due to hydro-gravitational interactions with the surrounding gas.  Usually, migratory torques are negative, leading to orbital inspiral, but sometimes they can become positive, leading to an outspiral.  When a zone of positive torque lies interior to one of negative torque, a migration trap forms \citep{PaardekooperMellema06}.  These traps can attract large numbers of stellar mass BHs into a narrow radial zone, facilitating close encounters and BBH formation \citep{McKernan+12, Bellovary+16, Secunda+19}.  The locations and even existence of these traps depend on the AGN disk structure \citep{GilbaumStone22, Grishin+2024}, as well as on radiative feedback from the embedded BH.  If migration traps exist, they can in principle catalyze many generations of hierarchical BBH mergers, but the extent of these merger trees depends on both ``input'' and ``output'' uncertainties.

On the {\it input} side, the rate at which individual BHs are delivered to the trap is uncertain, set by either the rate of {\it in situ} BH formation \citep{Levin07, GilbaumStone22} or the rate of capture due to gas drag.  Multi-body interactions within the trap may also be limited by the complex resonant chains that form under at least some circumstances \citep{Secunda+19, Secunda+20}.  In this paper, however, we quantify the {\it output} uncertainties, namely, the ways in which BHs may {\it permanently exit} from a trap.  Following a BBH merger, the resulting BH may be ejected due to GW recoil \citep{Fitchett83, Gonzalez+07}; or it may grow so large that its gravity opens a gap in the disk and eliminates the trap \citep{Ward97, Kanagawa+18}; or, most prosaically, the trap may go away at the end of the AGN lifetime.

In this paper, we quantify these different ``exit strategies,'' as well as their implications for both LVK-band GW signals and future EMRI GWs.  In \ref{sec:methods}, we present simple models for AGN disks, migratory torques, GW recoil kicks, and gas capture/alignment processes.  In \ref{sec:results}, we present the results of our 
calculations to estimate the statistical properties of BHs that eventually escape from migration traps.  We conclude in \ref{sec:discussion}.

\section{BBH Mergers in AGN}
\label{sec:methods}
We employ 
root-finding to solve the AGN 1D disk model presented by \cite{SirkoGoodman2003}, similar to the approach in \cite{Grishin+2024} (see App. \ref{app :disk}, where the relevant notation is defined). 
This model assumes axisymmetry, thermal stability, and vertically averaged profiles. At smaller radii, where the Toomre instability parameter \citep{Toomre1964} $Q_{\rm T}>1$, we utilize the canonical \cite{ShakuraSunyaev1973} solution, in which the central SMBH mass $M$, the AGN accretion rate $\dot{M}$, and the dimensionless viscosity $\alpha \le 1$ are model parameters. In regions where a Shakura-Sunyaev disk becomes Toomre unstable, we set $Q_T=1$. 
Our disk calculations incorporate opacity tables from \cite{IglesiasRogers1996} and \cite{Semenov+2003}.

As discussed in \S \ref{introduction}, migration traps are potentially crucial for BBH mergers (either hierarchical or single-generation). To locate these traps, we calculate the net torque exchange between an embedded  black hole and the surrounding gas. This calculation combines type I torques (gravitational perturbations due to resonant interactions) and thermal torques (perturbations arising from temperature gradients caused by hot/cold perturbers\footnote{We account for thermal torque saturation using eq. 36 in \citealt{Grishin+2024}.}), as detailed in \cite{JimenezMasset17}, \cite{Masset17}, \cite{Grishin+2024}, and App. \ref{app: torques}. While type I torques are consistently negative in AGN conditions \citep{Grishin+2024}, thermal torques can change signs. Migration traps/anti-traps form at locations where the net torque transitions from negative to positive (or vice versa).

\begin{figure}
    \centering
    \includegraphics[width=1\linewidth]{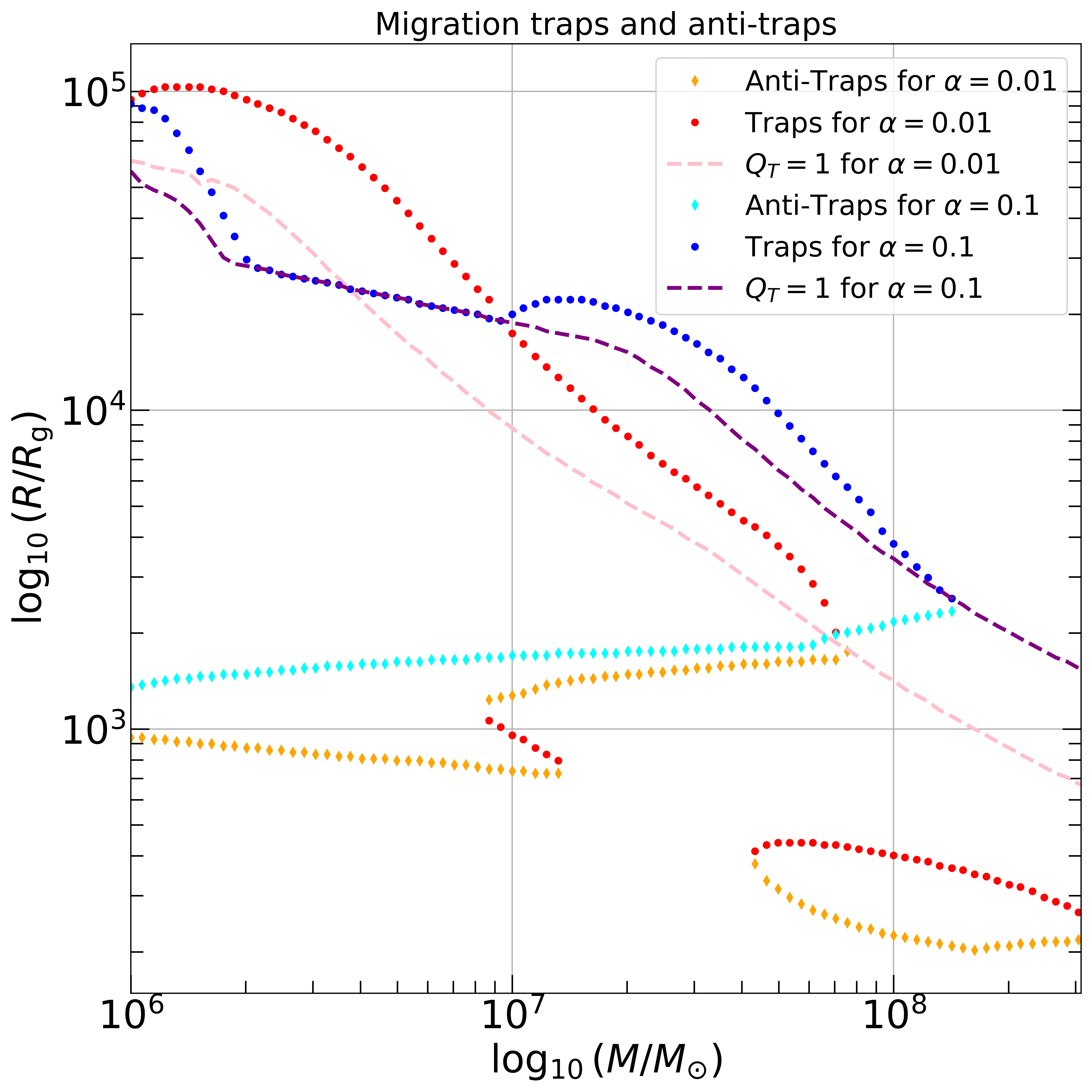}
    \caption{The radial locations $R_{\rm trap}$ of migration traps and anti-traps as a function of SMBH mass $M$, for Eddington ratio $\dot{M}/\dot{M}_{ \rm Edd} =0.1$ and the perturber BH mass  $m_{\bullet}=10 M_\odot$. \textcolor{blue}{Blue} (\textcolor{cyan}{cyan}) points represent the traps (anti-traps) for $\alpha = 0.1 $ and \textcolor{red}{red} (\textcolor{orange}{orange}) points represent $\alpha = 0.01$. The two dashed lines show the location where the Toomre parameter $Q_{\rm T}=1$ for each $\alpha$. For the $\alpha=0.01$ case there are some masses $M$ that produce two traps; at the highest $M$ values, no traps exist \citep{Grishin+2024}.}
    \label{fig: Traps}
\end{figure}

Figure \ref{fig: Traps} illustrates these calculations, displaying trap locations $R_{\rm trap}$ for Eddington ratio\footnote{Throughout this paper, we take the Eddington accretion rate $\dot{M}_{\rm Edd}=L_{\rm Edd}/(\eta c^2)$ with $\eta=0.1$ and $L_{\rm Edd}=3.77\times 10^4 L_\odot (M/M_\odot) $} $\dot{M}/\dot{M}_{\rm Edd}=0.1$ and $\alpha={0.01,0.1}$. In both cases, a primary trap/anti-trap pair is evident up to approximately $M\approx 10^8 M_\odot$, with $R_{\rm trap}$ near the $Q_{\rm T}=1$ radius.  The ``watershed'' (areas between the innermost anti-traps and outermost traps) in between becomes narrower and disappears for $M \gtrsim 10^8 M_\odot$, although an isolated ``island'' persists st somewhat larger masses when $\alpha$ is small. Increasing the perturber mass does not alter the locations of traps/anti-traps in these regions, because they are set by Eddington-limited, unsaturated thermal torques, which are mass-independent \citep{Grishin+2024}. However, the island regions decrease in area  (and eventually vanish) with increasing perturber mass, due to the thermal torque saturation correction (see App. \ref{app: torques}).  We note that the island traps arise from a delicate balance between (positive) thermal torques and (negative) Type I torques, and are more sensitive to the details of torque prescriptions than the primary traps are.

When the perturber is massive enough, it creates a gap in the disk \citep{Ward97}, negating the validity of type I and thermal torque calculations.  We employ a realistic\footnote{Note that much previous work \citep{McKernan+14,Stone+17,GilbaumStone22} used a gap opening criterion based on the older prescription of \cite{Crida+06}, which is more restrictive and less accurate.} criterion for the minimum BH mass required to open a gap \citep{Kanagawa+18,Tagawa+20}:
\begin{equation}
    m_\bullet \ge M_{\rm gap} \equiv M \sqrt{25 \alpha (H/R)^5}.
\end{equation}

For a stellar BH mass $m_\bullet$, the disk's steady state gas surface density $\Sigma$ will drop to a value $\Sigma_{\rm gap} \approx \Sigma / (1+ K/25)$ where $K=25(m_\bullet/M_{\rm gap})^2$ (see App. \ref{app: torques}). For large $K\gg 25$, the migration timescale becomes independent of the mass of the BH, as expected for type II migration, and the thermal torques vanish due to the greatly reduced optical depth $\tau \propto \Sigma_{\rm gap} \propto \Sigma/K$. Thus we expect only inward migration for very deep gaps.  
In \S\ref{sec:results} we discuss the implications of the softer criterion for gap opening. 

Once two single BHs arrive in the same migration trap, we can expect relatively rapid pairing \citep{Qian+2024}.
After a binary forms, it may merge either due to hydrodynamical forces or due to repeated encounters with third bodies that subsequently approach the trap.  Hydrodynamical forces will initially take a form akin to gas dynamical friction \citep{Baruteau+11}, but once the binary hardens sufficiently, further evolution will be mediated by a circumbinary flow \citep{LiLai22, LiLai23}.  While multi-dimensional hydrodynamical simulations find that this is a complex problem, two qualitative conclusions from the current literature \citep{Dempsey+22, LiLai24} are relevant: (i) prograde binaries may either contract or expand (so gas-driven coalesence is not guaranteed); (ii) orbital evolution occurs on timescales close to the mass-doubling timescale.  This can be short if circumbinary disks accrete at their Bondi-Hoyle rate \citep{Stone+17}, but will become the long Salpeter time ($\sim 10^7~{\rm yr}$) if the time-averaged accretion rate is Eddington-capped by feedback\footnote{If time-averaged accretion rates onto single BHs or BBHs are not Eddington-capped by feedback, a small number of embedded compact objects will be able to consume the large majority of AGN gas \citep{GilbaumStone22}.}.

Given these two potentially large problems with gas-assisted hardening, we estimate the binary-single merger time based on interactions with other BHs in the trap. Consider a BBH with equal BH masses $m_\bullet$ and separation $a$, at a distance $R$ from the SMBH. A single black hole of the same mass also orbits the SMBH at a distance $R+\delta$ (both orbits are prograde). Every time 
$P_{\rm fly} \approx (R/\delta) \Omega^{-1}$,a close encounter between the binary and the tertiary occurs with relative velocity $\Delta v\approx (1/2) \sqrt{GM/R} (\delta / R)$.  If $\delta < \delta_{\rm c} \approx \sqrt{6 G m_\bullet a / \Delta v^2}$, a critical value, then a strong binary-single scattering can ensue, which will decrease 
the separation of the binary\footnote{This is true in a statistical sense so long as the binary is ``hard'' (i.e. \citealt{Heggie75}).  We show in Appendix \ref{app : timescales} that this is generally the case. 
}. 

If both orbits are sufficiently circular and coplanar, then $\delta < \delta_{\rm c}$ is a sufficient condition to produce strong binary-single scatterings after each time $P_{\rm fly}$; conversely, if (e.g.) the outer tertiary has a strongly eccentric or inclined orbit, then it may take many flyby times for a strong scattering to result, as the binary and tertiary will typically be far apart in radius or vertical height during each azimuthal alignment.  As the binary hardens, $\delta_{\rm c}$ shrinks, and so an initially ``deterministic'' binary-single pair may transition into the less favorable ``stochastic'' regime for producing strong encounters.  While a full derivation of the relevant dynamics (including a generalization to unequal binary-single masses) is given in Appendix \ref{app : 2+1 merger}, we give here the timescale for equal-mass binary-single systems to undergo a strong encounter, $T_{\rm s}$, specializing to the more relevant gravitationally focused limit $T_{\rm s}^{\rm f}$ of binary-single scatterings.  In the deterministic (large $a$, large $\delta_{\rm c}$) limit, 
\begin{equation}
T_{\rm s, d}^{\rm f} = P \left(\frac{2^{2/3}}{3\tilde{a}} \right)^{1/4} \left(\frac{m_\bullet}{M} \right)^{-1/3},
\end{equation}
while in the stochastic (small $a$, small $\delta_{\rm c}$) limit,
\begin{equation}
T_{\rm s, s}^{\rm f} = P \left(\frac{1}{3\times 2^{4/3}\tilde{a}} \right)^{1/2} \left(\frac{m_\bullet}{M} \right)^{-2/3} \sin(i). \label{eq:Tsf}
\end{equation}
In both these equations, we take $\delta=\delta_{\rm c}$, use the local orbital period of the binary $P(R)$, and write a dimensionless $\tilde{a}=a/R_{\rm t}\le 1$, where the tidal radius $R_{\rm t}=R(2m_\bullet / M)^{1/3}$.  In Eq. \ref{eq:Tsf}, we assume an orbital inclination for the tertiary, $i$, that satisfies $\delta / R \ll \sin(i) \ll 1$.
For practical evaluations, we set $\sin(i)=H/R$ as a reasonable upper limit.

Each strong encounter reduces the binary semimajor axis by a fraction $f_{\rm s} \approx 0.3$ \citep{StoneLeigh2019}. Eventually, GW emission will take over and drive the binary to merger when $T_{\rm GW}(a_{\rm GW}) \leq T_{\rm s}$, where the GW emission timescale $T_{\rm GW}$ is given by \cite{Peters1964}, and $a_{\rm GW}$ is the separation at which the two timescales are equal. As $T_{\rm s}$ grows with shrinking $a$, the final merger time from any initial separation reduces to a few GW times at $a=a_{\rm GW}$ (see Appendix \ref{app : 2+1 merger}):
\begin{equation}
T_{\rm m}  \approx \frac{1}{f_{\rm s}}T_{\rm  GW}\left(a_{\rm GW}\right) \label{eq: 2+1 merger time}.
\end{equation}
This is ultimately a conservative estimate, as the high eccentricities sometimes achieved in the aftermath of a strong binary-single encounter \citep{StoneLeigh2019} can greatly hasten a GW inspiral \citep{Peters1964}, and initial eccentricties and inclinations  of the binary can vary the merger rates as well \citep{Trani+2024}.

The calculations above describe the interaction with a single BH at distance $\delta$. However, as multiple BHs may in principle exist near migration traps, we also consider the effect of $N_\bullet$ BHs within a ring of width $\delta$. In this scenario, the interaction time $T_{\rm s}$ is on average reduced by a factor of $N_\bullet$. 
We constrain $N_\bullet$ based on two considerations. First, the total mass of embedded objects must not exceed the mass of gas in the region:
$N_\bullet m_\bullet \leq 2 \pi R  \Sigma \delta$. Second, the BHs must not be so tightly distributed in the annulus that the average distance between them becomes smaller than the wavelength of type I density waves: $2\pi R/N_\bullet \geq H$. Such a configuration would negate any torque exchange between the gas and BHs. 
 Consequently, the maximum number of BHs in the trap region is:
\begin{equation}
N_{\bullet,{\rm max}} = \left [ \frac{2 \pi R}{H} \min \left(1, \frac{\delta H \Sigma}{m_\bullet}  \right)
\right ].
\end{equation}
In later calculations of binary-single hardening, we compute both a minimum merger time ($N_\bullet = N_{\bullet, \rm max}$) and a maximum merger time ($N_\bullet = 1$).

An additional timescale of relevance is the rate at which the disk can supply BHs to the trap region. We characterize this supply rate as $\dot{N}_{\bullet} = f_{\rm N} \dot{M}/ \left< m_\bullet \right>$, where $f_{\rm N}$ is the ratio of the BH migratory mass flux to the steady state mass accretion rate, and $\left< m_\bullet \right>$ is the typical black hole mass. The corresponding timescale for black hole replenishment is thus $T_{\rm supply} = 1/\dot{N}_{\bullet}$. $f_{\rm N} \ll 1$ must remain quite small, as otherwise the embedded BHs would accrete more gas than the central SMBH, violating our assumptions about the disk structure and dynamics.

Post-merger, the merged BH may experience a strong recoil kick 
with a magnitude and direction that depend on mass ratio\footnote{Because general relativity is scale-free, the absolute masses of the BBH components do not affect the kick (see Appendix \ref{app:recoils}).} $q\le 1$, spin directions, spin magnitudes, and orbital phase angles. With typical kick velocities $v_{\rm k} \sim 10^{2-3}~{\rm km~s}^{-1}$, these recoils can eject merger products from many astrophysical environments \citep{Campanelli+2007,Gonzalez+07,LoustoZlochower2011,Rodriguez+2019}.  We calculate the statistical properties of BBH merger products in AGN disks, assuming the embedded BBH's orbital angular momentum is aligned with that of the AGN, and slightly misaligned spins (by up to $30^{\circ})$. 
We simulate merger product kicks using post-Newtonian fitting formulae from \citet{Gerosa+2023}\footnote{https://github.com/dgerosa/precession}. The simulated kick velocity is added to the existing Keplerian velocity (assuming a circular orbit around the SMBH), yielding post-merger orbital parameters such as inclination $i$, eccentricity $e$, and semi-major axis $a$. We repeat this process, randomly sampling phase angles $10^4$ times at each  value of $R/R_{\rm g}$.  
 
To permit hierarchical mergers in the migration trap, the merger product must remain bound; it must also either stay in the trap region or return faster than the AGN lifetime. When a recoil kick increases orbital inclination above the disk plane ($i>\tan ^{-1} (H/R)$) without full ejection, we calculate the realignment time $T_{\rm RA}$ using the formulation of \cite{GenerozovPerets2023} that considers both gas dynamical friction and aerodynamic gas drag (see Appendix \ref{app : timescales} and \citealp{gri15}). While lifetimes of individual AGN episodes are quite uncertain, we compare to an upper limit for AGN lifetimes calculated using observational data from \cite{Aird+2018} and methods described in \cite{GilbaumStone22} and \cite{HopkinsHernquist2009} (Appendix \ref{app : timescales}).

To summarize, we identify migration trap locations $R_{\rm trap}$ for various disk parameters (varying mass $M$, Eddington ratio $\dot{M}/\dot{M}_{\rm Edd}$, and viscosity $\alpha$; see Fig. \ref{fig: Traps}). At these traps, we evaluate the statistics of post-merger kicks and their aftermath. We compute the gap opening mass and all relevant timescales (AGN lifetime, migration, realignment, binary-single merger) to assess what ultimately limits hierarchical mass growth in migration traps.

\section{Results}
\label{sec:results}

\begin{figure}
    \centering
    \includegraphics[width=1\linewidth]{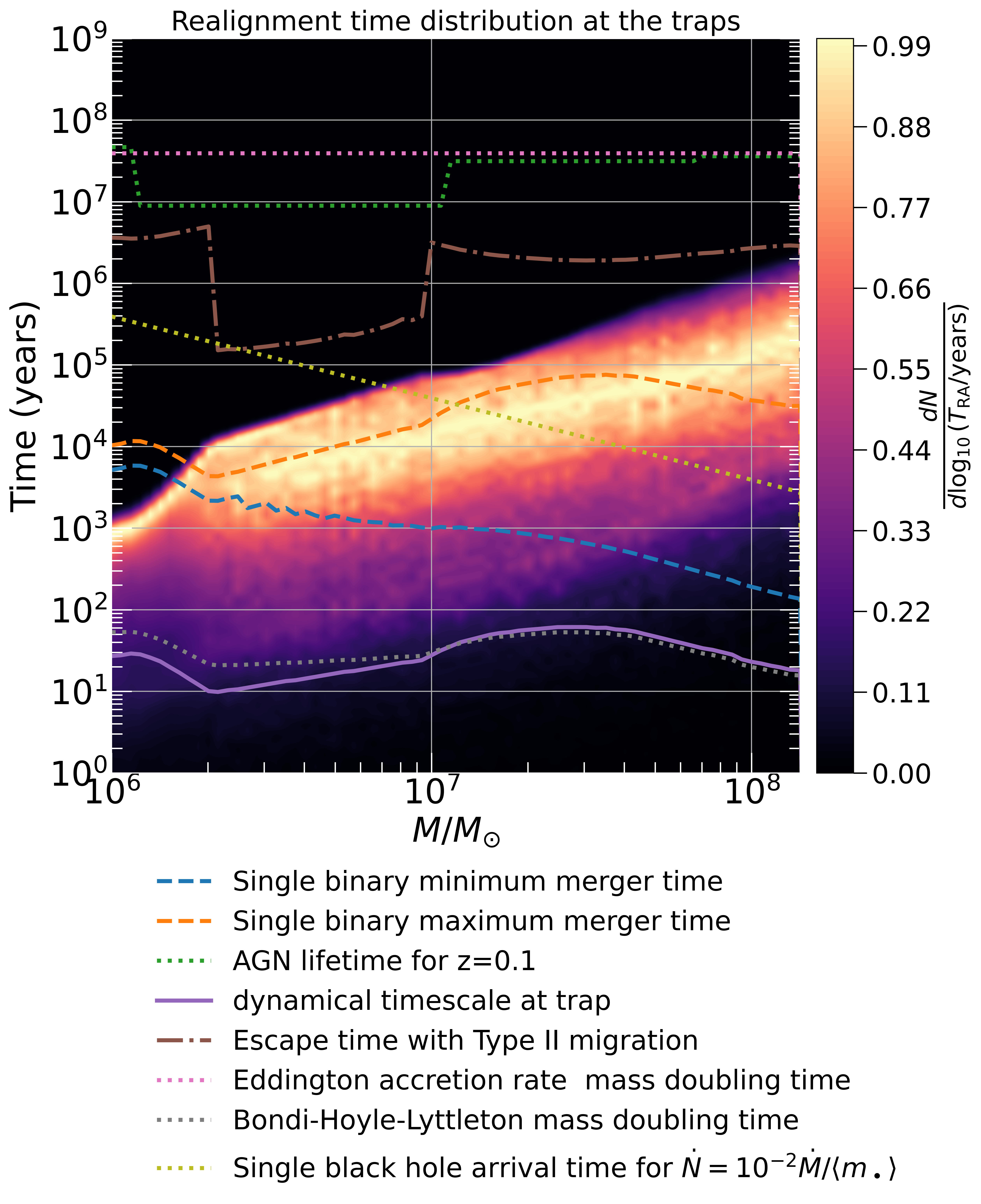}
    \caption{Normalized distributions of realignment times $T_{\rm RA}$ for a $m_\bullet = 10 M_\odot$ object in the outermost migration trap, shown as a function of SMBH mass $M$, fixing the AGN Eddington ratio $\dot{M}/\dot{M}_{\rm Edd} =0.1$ and $\alpha = 0.1$. Color-coding represents the probability density for different $T_{\rm RA}$ assuming a mass ratio $q=1$, spin magnitude of $\chi_{1,2}=0.7$ for both black holes, and random spin misalignment of up to $30^{\circ}$ with respect to the angular momentum of the AGN disk. The \textcolor{green}{green} dotted line represents upper limits on observed AGN lifetimes (App. \ref{app : timescales})  at redshift $z=0.1$.  The \textcolor{orange}{orange} and \textcolor{blue}{blue}  dashed lines represent the approximate maximum (minimum) times for a merger, due to $2+1$ scattering against 1 ($N_{\rm max}$) BHs. The \textcolor{violet}{violet} solid line represents the local dynamical time at the trap.  The \textcolor{pink}{pink} (\textcolor{gray}{gray})  dotted line is the mass doubling time of an embedded object  via Eddington (Bondi-Hoyle-Lyttleton) rate accretion.  The  dash-dotted \textcolor{brown}{brown}  shows the time for a sufficiently heavy BH to escape the entire trap region via Type II migration. The \textcolor{yellow}{yellow} dotted line represents an example timescale for the influx of black holes to the trap ($1/\dot{N}$).}
    \label{fig: realignment_time}
\end{figure}

In agreement with our past work \citep{Grishin+2024}, we find that up-to-date calibrations \citep{JimenezMasset17} of Type I torques are negative-definite in realistic AGN disk models, and cannot produce migration traps on their own.  High Bondi-Hoyle accretion rates do facilitate positive thermal torques \citep{Masset17}, however, generating traps in AGN below a critical luminosity threshold $L_{\rm AGN} \sim 10^{45}~{\rm erg~s}^{-1}$ (the precise threshold depends on $\alpha$ and other disk microphysics). 

We have computed the realignment time $T_{\rm RA}$ for a broad variety of AGN parameters and pre-merger BBH properties. In almost all relevant cases,
BBH mergers in migration traps generate insufficient GW recoil to escape the system entirely (see App. \ref{app:recoils}), though excitation to inclined orbits is common.  Fig. \ref{fig: realignment_time} shows a representative example of $T_{\rm RA}$ distributions for typical AGN parameters ($\dot{M}/\dot{M}_{\rm Edd} = 0.1$, $\alpha=0.1$).  Typical realignment times range from $\sim 10$ to $\sim 10^{6}~{\rm yr}$, which we show in the figure is far shorter than effective AGN lifetimes across cosmic history \citep{HopkinsHernquist2009}.  The lifetime of individual AGN episodes is shorter than the effective AGN lifetime of a galaxy, but is still likely much longer than this range of $T_{\rm RA}$.  

In Fig.~\ref{fig: realignment_time} we compare minimum and maximum times for mergers driven by binary-single scatterings and gravitational-wave emission (eq. \ref{eq: 2+1 merger time}). 
We can see that the orbital realignment times are comparable to or smaller than the upper limit of the scattering merger time. Coupled with the rapid dynamically-driven pairing by \cite{Qian+2024}, this finding suggests that realignment will not seriously delay migration trap mergers. 

Since the realignment of recoil-tilted orbits is unlikely to be the rate-limiting step in hierarchical BH growth, we consider other factors. In Fig. \ref{fig: realignment_time} we also show the black hole supply timescale $T_{\rm supply} = 1/\dot{N}_{\bullet}$,  which is only comparable to the merger timescales for the very optimistic case of $f_{\rm N} = 0.01$. For more realistic, lower values of $f_{\rm N}$, the rate at which the disk can supply black holes to the trap will grow and may become the limiting factor in determining the overall merger rate. This suggests that even in scenarios where orbital realignment and binary-single interactions are efficient, the trap's ability to facilitate hierarchical growth may ultimately be limited by the availability of new black holes.

\begin{figure}
    \centering
    \includegraphics[width=1\linewidth]{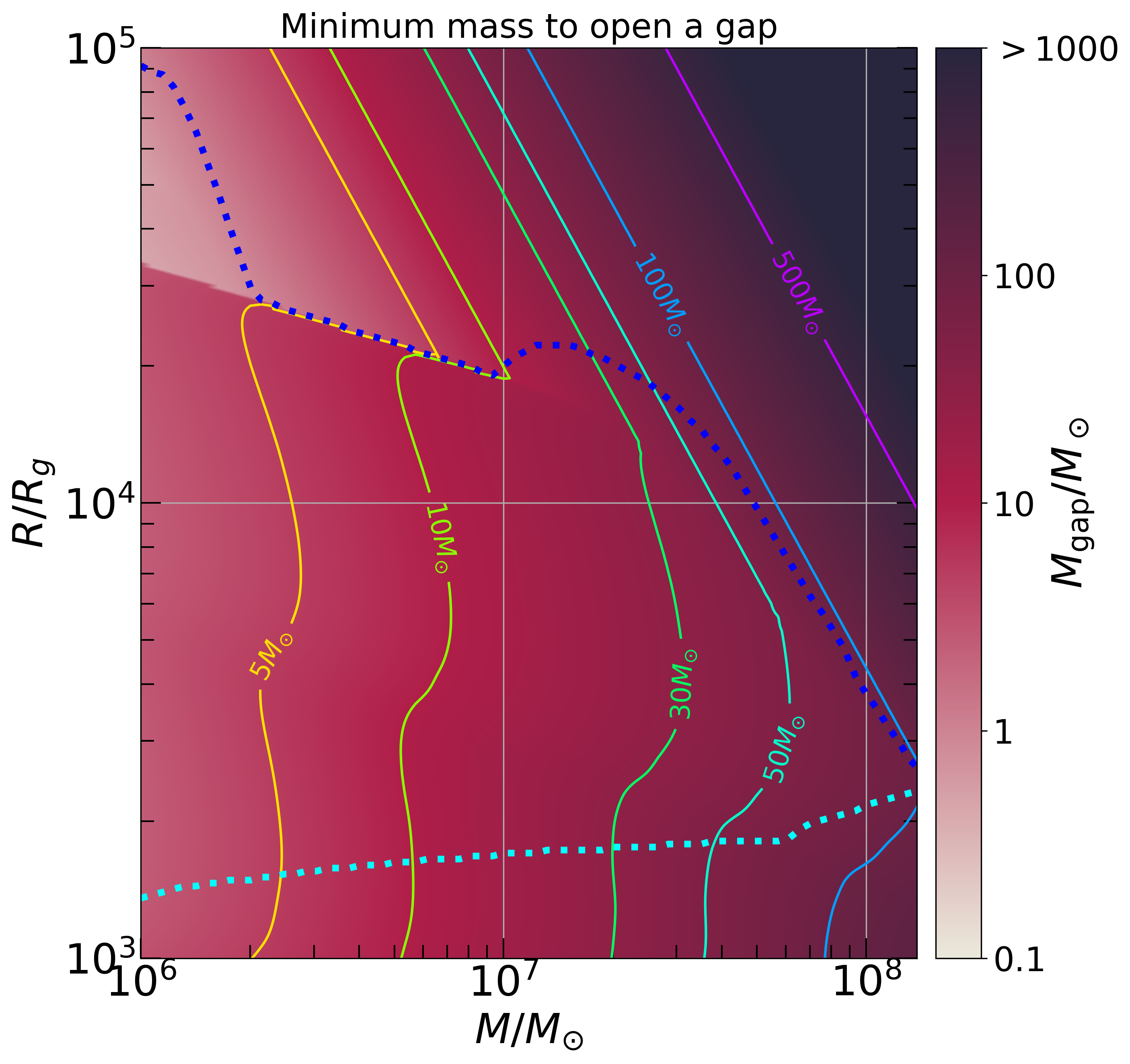}
    \caption{The color-coded minimum mass for opening a gap, $M_{\rm gap}$, plotted in the 2D parameter space of SMBH mass $M$ and radial location $R/R_{\rm g}$. This plot takes the Eddington ratio $\dot{M}/\dot{M}_{\rm Edd} =0.1$ and $\alpha = 0.1$. Labeled contours highlight particular values of $M_{\rm gap}$. The \textcolor{blue}{blue} (\textcolor{cyan}{cyan}) dotted lines indicate the trap (anti-trap) locations. At low SMBH masses, even small stellar mass BHs can open gaps at all radii, making hierarchical growth unlikely in migration traps.}
    \label{fig: mass_for_gap}
\end{figure}

Next  we consider gap opening as a potential limiting factor.  In Fig. \ref{fig: mass_for_gap}, we show the minimum mass for an embedded BH to open a gap in the AGN disk with $\dot{M}/\dot{M}_{\rm Edd}=0.1$ and $\alpha=0.1$.  The updated gap-opening prescription of \citet{Kanagawa+18} predicts relatively low gap-opening masses $M_{\rm gap} \sim 5-10 M_\odot$ for smaller SMBHs ($M \sim 10^{6-7}M_\odot$).  For heavier SMBHs, the gap-opening mass can reach $\sim 100 M_\odot$, but rarely larger than this at $R=R_{\rm trap}$.  Gap-opening masses can be somewhat larger for denser AGN gas (e.g. $\alpha=0.01$; Fig. \ref{fig: mass_for_gap, alpha =0.01}), but in general, gap-opening appears to be the most robust way for BHs to escape from migration traps.  Indeed, gap-opening is such an efficient escape mechanism that in many cases it may prevent hierarchical growth altogether, at least for $M \lesssim 10^7 M_\odot$.

Once a BH opens a gap at the trap location, it will migrate inwards in a Type II fashion.  While $M_{\rm gap}$ is usually not a strong function of radius in our AGN models, under some circumstances it may increase inwards (see e.g. $10^7 \lesssim M/M_\odot \lesssim 10^8$ regions in Fig. \ref{fig: mass_for_gap, alpha =0.01}), in which case the gap will eventually close and the BH may become stuck (provided it is still outside the thermal anti-trap).  This new type of ``gap-closing trap'' may eventually go away as more BHs arrive, merge, and accumulate enough mass to open a gap once again, although the dynamics of this situation are complex and merit further investigation.  Once the BH migrates inwards of the thermal anti-trap, it will become a ``wet EMRI,'' inspiraling into the central SMBH and producing a {\it LISA}-band GW signal \citep{Levin07, DerdzinskiMayer23}.  While migration traps could in principle inhibit wet EMRI production, the low gap-opening masses of smaller SMBHs suggest that this will not be a strong barrier in practice.

\begin{figure}
    \centering
    \includegraphics[width=1\linewidth]{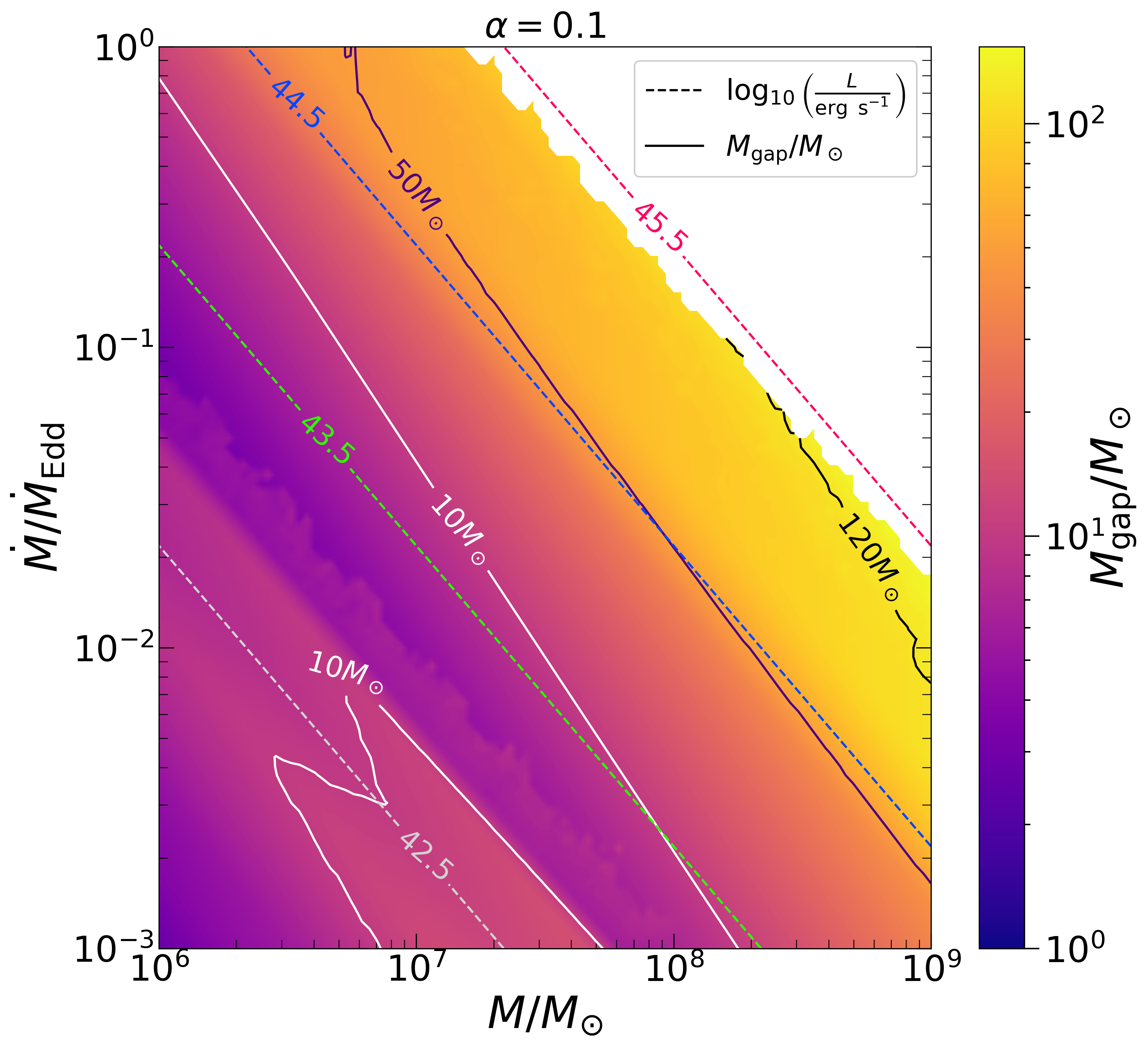}
    \caption{ The minimum mass ($M_{\rm gap})$ for escaping the migration trap via gap opening for an $\alpha=0.1$ disk, shown in color coding as a function of SMBH mass $M$ and AGN Eddington ratio $\dot{M}/\dot{M}_{\rm Edd}$. The solid contour lines highlight individual values of $M_{\rm gap}$ as labeled. The dashed contour lines represent the AGN luminosity $L$, assuming a radiative efficiency $\eta=0.1$.  Hierarchical growth in traps only occurs in a relatively narrow range of AGN luminosities, for $10^{43.5}~{\rm erg~s}^{-1} \lesssim L \lesssim 10^{45.5}~{\rm erg~s}^{-1}$.  At higher luminosities, the traps fail to exist \citep{Grishin+2024}; at lower luminosities, gaps open too easily to retain merger products in the traps. The maximum value we find is $M_{\rm gap} \approx 150 M_\odot$, though larger masses arise for lower $\alpha$ (see App. \ref{app : additional results} and Fig. \ref{fig: mass_for_escape, alpha =0.01}).} 

    \label{fig: mass_for_escape}
\end{figure}

In Fig. \ref{fig: mass_for_escape}, we show the relationship between $M_{\rm gap}$ and total AGN luminosity $L_{\rm AGN}=\eta \dot{M} c^2$ (we assume a radiative efficiency $\eta=0.1$ throughout), for the same AGN models as in Figs. \ref{fig: realignment_time} and \ref{fig: mass_for_gap}.  As in \citet{Grishin+2024}, there is a maximum $L_{\rm AGN} \sim 10^{45}~{\rm erg~s}^{-1}$ above which traps stop existing.  However, we now also find a {\it minimum} $L_{\rm AGN}$ below which $M_{\rm gap}$ is so small that hierarchical mergers/growth are essentially impossible.  There is ultimately a narrow range of luminosities, $10^{43.5}~{\rm erg~s}^{-1} \lesssim L_{\rm AGN}  \lesssim 10^{45.5}~{\rm erg~s}^{-1}$, for which hierarchical growth is possible at all.  For hierarchical growth into the pair instability mass gap, this range is even narrower, $10^{44.5}~{\rm erg~s}^{-1} \lesssim L_{\rm AGN}  \lesssim 10^{45.5}~{\rm erg~s}^{-1}$.  This conclusion depends weakly on the underlying AGN parameters: see Fig. \ref{fig: mass_for_escape, alpha =0.01} for an alternate $\alpha=0.01$ case.  In contrast, $M_{\rm gap}$ can be more sensitive to the AGN parameters.  Comparing Figs. \ref{fig: mass_for_escape} and \ref{fig: mass_for_escape, alpha =0.01}, we see that $M_{\rm gap}$ is generally similar, but for $M \gtrsim 10^8 M_\odot$ and low $\dot{M}/\dot{M}_{\rm Edd}$, low-$\alpha$ AGN disks can achieve much higher values of $M_{\rm gap}$, in some cases reaching $M_{\rm gap}\sim 1000 M_\odot$.

\section{Discussion and Conclusions}
\label{sec:discussion}

In this paper, we have explored different ways for hierarchical growth to come to terminate for BHs in AGN migration traps.  Our primary conclusions are as follows:
\begin{enumerate}
\item  GW recoil kicks are almost never able to unbind merger products from the AGN.  While BHs can often be kicked out of the disk ($i > H/R$), realignment times are fast, and ultimately kicks do not prevent repeated mergers.
\item Likewise, merger times for BBHs in migration traps are relatively short.  Except in very short-lived AGN episodes (lifetimes $<10^6$ yr), it is unlikely that a BH in a migration trap can ``wait out'' a sequence of hierarchical mergers.
\item  Gaps in AGN disks open quite easily, especially at low SMBH masses.  Gap-opening represents the most likely escape route for BHs in migration traps, and indeed, may prevent the practical existence of traps altogether for SMBHs with $M \lesssim 10^7 M_\odot$.  Because this is the mass range relevant for {\it LISA}, wet EMRI masses will not be boosted by hierarchical growth in traps.
\item For higher-mass SMBHs ($M \gtrsim 10^7 M_\odot$), hierarchical growth in the traps occurs until a gap opens.  This allows the merger product to migrate inwards, although sometimes a "gap-closing" trap can appear at smaller radii, leading to further growth.

\item Our timescale analysis shows that migration traps may often be empty: when the BH supply rate $\dot{N}_\bullet$  is low, and merger times are comparatively short, BHs merge soon after reaching the trap. The merger products then escape via gap opening, preventing accumulation of BHs in the trap region. In this regime, the trap acts more as a merger site than a growth site, with its efficiency limited by the rate at which the disk can supply new BHs.

\item Traps can only produce pair instability mass-gap LVK mergers in a finite band of AGN luminosities, with a minimum $L_{\rm AGN} \sim 10^{44.5}~{\rm erg~s}^{-1}$ and a maximum $L_{\rm AGN}\sim 10^{45.5}~{\rm erg~s}^{-1}$.  This may facilitate statistical tests \citep{Bartos+17b, Veronesi+23} of the AGN channel by narrowing down the permitted host AGN for the high-mass subset of LVK signals.
\end{enumerate}

Several aspects of our calculations should be examined more carefully in future work.  Some results (e.g. $R_{\rm trap}$) depend on disk model, particularly for the $Q_{\rm T}<1$ regions. Our treatment of Type II torques assumes they are negative-definite, and our proposed gap-closing traps represent a limit of Type II migration, as well as thermal torques,  that has not previously been explored with hydrodynamical simulations. These areas of Type II migration and thermal torques should be studied with further simulations.  In computing binary-single hardening times, we have neglected resonant chains \citep{Secunda+19}, which can play an important role in modifying these dynamics.

Several remaining questions contribute to the largest uncertainties in the AGN channel for BBH formation. First, as shown in our timescale analysis, the (uncertain) rate of black hole influx into the trap region may be the rate-limiting step that determines the total merger rate, particularly for realistic (low) values of the mass flux fraction $f_{\rm N}$. Additionally, the role of accretion feedback from embedded BHs remains unclear, as such feedback may modify the overall structure of the AGN disk \citep{GilbaumStone22}. Calculation of thermal torques is sensitive to BH luminosity; as in past work \citep{GilbaumStone22, Grishin+2024}, we have assumed that accretion onto embedded BHs is (at least in a time-averaged sense) Eddington-capped, but this must be checked with radiation-hydrodynamics simulations.

\section*{Acknowledgements}

This research was partially funded by the Australian Research Council Centre of Excellence for Gravitational Wave Discovery (OzGrav), through project number CE230100016; and by ARC Future Fellowship FT190100574 to IM.  NCS and SG gratefully acknowledge support from the Israel Science Foundation (Individual Research Grant 2565/19) and the Binational Science Foundation (grant Nos. 2019772 and 2020397).

\bibliographystyle{aasjournal}
\bibliography{main}

\appendix

\section{Methodology}
\label{app: methodology}

Here we provide precise details of the models employed in this work for the hydrodynamic structure of the AGN disk, the recoil kicks imparted by anisotropic GW emission, the torques that lead embedded BHs to migrate through an AGN disk, and the hydrodynamic process of realignment (for BHs kicked to inclined orbits).

\subsection{AGN Disk Model}
\label{app :disk}
In this work, using a modified version of PAGN code \citep{Gangardt+2024}\footnote{https://github.com/DariaGangardt/pAGN}, we solve the disk model in \cite{SirkoGoodman2003} using its continuous branch of solutions.  This is a 1D (i.e axisymmetric, vertically averaged), steady state thin disk model similar in spirit to that of \citet{ShakuraSunyaev1973}. The two differences from the classic work of \citet{ShakuraSunyaev1973} are (i) the use of realistic, tabulated opacities \citep{IglesiasRogers1996, Semenov+2003} and (ii) a different physical treatment of Toomre unstable zones. 
Specifically, \cite{SirkoGoodman2003} assume that in regions where the gas is cold enough for clumps to form (or in other words, for the Toomre parameter $Q_T<1$; \citealt{Toomre1964}), there will be star formation or some other feedback mechanism which will heat the disk until the the Toomre parameter self-regulates to a marginally stable $Q_{\rm T}=1$ value.  Angular momentum transport is assumed to be driven by a local effective viscosity parameterized by the usual dimensionless $0<\alpha<1$ \citep{ShakuraSunyaev1973}.

Here we list the relevant AGN disk parameters and variables: \begin{align*}
Q_{\rm T}  & - \quad \quad \quad\text{Toomre instability parameter; if $Q_{\rm T}<1$, gas is unstable } \\
 & \quad \quad \quad \quad \text{to density perturbations and stars may form} \\
M & - \quad \quad \quad \text{SMBH mass} \\
\epsilon_{\rm Edd} &  - \quad \quad \quad\text{Eddington ratio of the disk's luminosity (and accretion rate)} \\
R & - \quad \quad \quad\text{Distance from the center of the disk} \\
\Omega & - \quad \quad \quad\text{Keplerian angular frequency $\left( \sqrt{GM/R^3} \right)$} \\
T & - \quad \quad \quad \text{Local central temperature of the disk} \\
c_{\rm s} & - \quad \quad \quad \text{Local sound speed} \\
\Sigma & - \quad \quad \quad \text{Local gas surface mass density} \\
\rho & - \quad \quad \quad \text{Local gas volume density} \\
H & - \quad \quad \quad\text{Local disk scale height} \\
P &  - \quad \quad \quad\text{Local gas total pressure} \\
\tau & - \quad \quad \quad \text{Local disk optical depth} \\
\nu & - \quad \quad \quad \text{Local gas effective viscosity} \\
\alpha  & - \quad \quad \quad  \text{Unitless Shakura-Sunyaev viscosity constant}
\end{align*}

\subsection{Torques and migration}
A stellar mass object of mass $m_\bullet$ embedded in an AGN disk will migrate (inwards or outwards) due to external forces acting on it.  A relatively simple, but astrophysically important, example of such a force is GW emission from the quadrupole moment formed with the MBH.  We include the standard, leading-order formulation of GW dissipation \citep{Peters1964} in our calculations, and outline here the more complicated gravitational-hydrodynamic drag forces at play.
\label{app: torques}
\subsubsection{Type I torques}
Type I torques are created due to the gravity of a perturber (i.e. the embedded BH) acting on the surrounding gas \citep{GoldreichTremaine80}. The gas over/under-densities produced by these gravitational perturbations produce a net torque acting on the perturber's orbit:
\begin{equation}
        \Gamma_{\rm I} = C_{\rm I} \frac{H}{R} \Gamma_0,
\end{equation}
which for a thin disk is a small fraction of the ``one-sided'' torque:
\begin{equation}
        \Gamma_0 = q_\bullet^{2}\Sigma R^{4}\Omega^{2} \left( \frac{H}{R} \right)^{-3}.  \label{eq: gamma_0}
\end{equation}
Here $q_\bullet= m_\bullet / M \ll 1$ is the mass ratio between the embedded object and the central MBH.  The dimensionless prefactor $C_{\rm I}$ has been calibrated by a number of hydrodynamic calculations and simulations; in this paper, we employ the recent calibration of \cite{JimenezMasset17}:
\begin{align}
    C_{\rm I} &= C_{{\rm L}}+(0.46- 0.96\nabla_{\Sigma}+1.8\nabla_{T})/ \gamma\\
    C_{\rm L} &= (-2.34+0.1\nabla_{\Sigma}-1.5\nabla_{T})f(\chi/\chi_c) \\
    f(x) & = \frac{(x/2)^{1/2}+1/\gamma}{(x/2)^{1/2}+1} \ .  
\end{align}
Here $\chi_c = H^2 \Omega^2$  and $\chi$ is the thermal diffusivity of the AGN gas \cite{Paardekooper2014}:
\begin{equation}
        \chi  =  \frac{16(\gamma -1) \sigma_{\rm SB} T^4}{3  \kappa \rho^2 H^2 \Omega^2  } \label{eq: chi}.
\end{equation}
Here we define dimensionless logarithmic derivatives of fluid quantities $X$ as $\nabla_X \equiv {\rm d}\ln X/{\rm d}\ln R$, and use a gas adiabatic index $\gamma$ (we take the fiducial value of $\gamma=5/3$).

\subsubsection{Thermal torques}
\cite{Masset17} has shown that the diffusion of heat released by an embedded perturber can also create linear density perturbations that likewise result in net torque exchange between the perturber and the surrounding gas.  The resulting ``thermal torque'' is
\begin{equation}
    \Gamma_{\rm th} =\Gamma_{\rm hot} +\Gamma_{\rm cold} =
    1.61 \frac{\gamma-1}{\gamma}\frac{x_c}{\lambda_{\rm c}}\left(\frac{L}{L_{\rm c}} -1\right) \Gamma_0 \ .
\end{equation}
Here $L$ is the emitted luminosity of the perturber, which for practical purposes we assume to be the perturber's Eddington luminosity $L_{\rm Edd} \approx 3 \times 10^4 L_\odot (m_\bullet/M_\odot)$, as is usually the case for stellar mass BHs embedded in AGN \citep{Stone+17, GilbaumStone22}.  We note that  $\Gamma_{\rm hot}/\Gamma_{\rm cold}=-L/L_c$.  The other parameters are defined by:
\begin{align}
    L_{\rm c} &= \frac{4\pi G m_\bullet \rho} \gamma{\chi} \\
    x_c &= \frac{\nabla_P c_s^2}{3\gamma R\Omega^2} \\
    \lambda_{\rm c} &=\sqrt{\frac{2\chi} {3\gamma \Omega}} \ .
\end{align}

Here $L_{\rm c}$ is a critical luminosity for the embedded BH that separates positive thermal torques from negative ones; $x_{\rm c}$ is a typical offset between the BH and corotating gas; $\lambda_{\rm c}$ is a typical size of the density perturbation caused by thermal effects.

We also add a correction related to thermal torque saturation as discussed in \citep[][Section 5.3]{Grishin+2024}: 
\begin{align}
    \mu_{\rm th} & = \frac{\chi}{c_s \times \min \{R_{\rm BH}, H  \}}<1 
    \\
    \Gamma_{\rm th}' & =  \Gamma_{\rm hot} \frac{4\mu_{\rm th}}{1+4\mu_{\rm th}} + \Gamma_{\rm cold} \frac{2\mu_{\rm th}}{1+2\mu_{\rm th}} \ .
\end{align}

\subsubsection{Type II torques}
If a perturber is  massive enough, a gap will open in the gas at the annulus of the perturber \citep{Ward97}, which can lead to dramatic changes in migration rates for BHs emebedded in AGN disks \citep{GoodmanTan04,McKernan+14,Stone+17,GilbaumStone22}. This gap has sometimes been understood to facilitate a qualitatively different type of torque exchange between the gas and the perturber than the linear type I case.  More recently it has been shown that gap opening is more of a continuous transition, and that torque exchange in a gap region is similar to type I torques, albeit with a correction that takes into account the lower but non-zero density of gas in the gap \citep{Kanagawa+18,Tagawa+20}:
\begin{align}
         \Gamma_{\rm II} &= \frac{\Gamma_{\rm L}+\Gamma_{\rm C} \exp \left(-K / 20\right)}{1+0.04 K} \\ 
        K & = q_\bullet^2 \alpha^{-1} \left(\frac{H}{R} \right)^{-5} \ .
\end{align}
 
Here $\Gamma_{\rm L}$ and $\Gamma_{\rm C}$ are Lindblad and co-rotationg resonance torques respectively. In \cite{Kanagawa+18}  they are very similar to the ones in \cite{JimenezMasset17}, proportional to the type I torques with some prefactors depedent on the temperature and surface density power laws.
For massive enough objects, the gap isolates that black hole from the environment and makes the thermal torque vanish. This results in the black hole eventually escaping type I migration traps.

\subsection{Recoil statistics and orbital parameters }
\label{app:recoils}

We want to quantify the outcomes of GW recoil kicks following BH-BH mergers in the plane of an AGN disk, so as to understand possible scenarios for hierarchical mergers.  
To compute GW kick velocity vectors, we use the calculation from \cite{Gerosa+2023}, in particular employing their existing code\footnote{https://github.com/dgerosa/precession}.  The computation of GW kick magnitudes and directions depends on the mass ratio $q\le 1$ between the two BHs, their dimensionless spin magnitudes ($\chi_1 \le 1$ and $\chi_2 \le 1$), their spin directions, and an orbital phase variable.  While we make astrophysically motivated assumptions for $q$, $\chi_{1,2}$, and the spin-orbit misalignment angle, this still leaves two\footnote{Both the orbital phase angle and the difference in azimuthal orientation of the two spin vectors are effectively isotropic.} angular variables that are essentially arbitrary.  We therefore must employ Monte Carlo methods to gain a statistical picture of the GW kicks imparted by BH-BH mergers; we do so by sampling these two azimuthal angles randomly and independently from $0$ to $2\pi$.

\begin{figure}
    \gridline{\fig{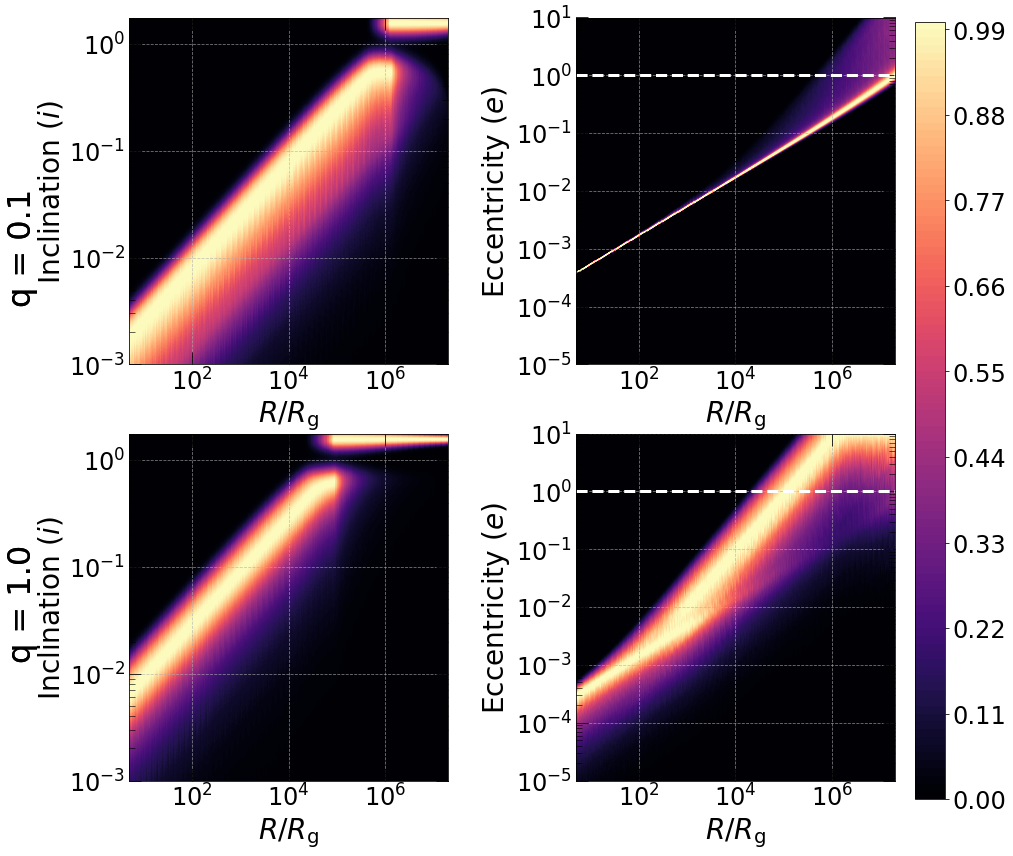}{0.49\textwidth}{(a) $\chi_{1,2}=0.7$}
             \fig{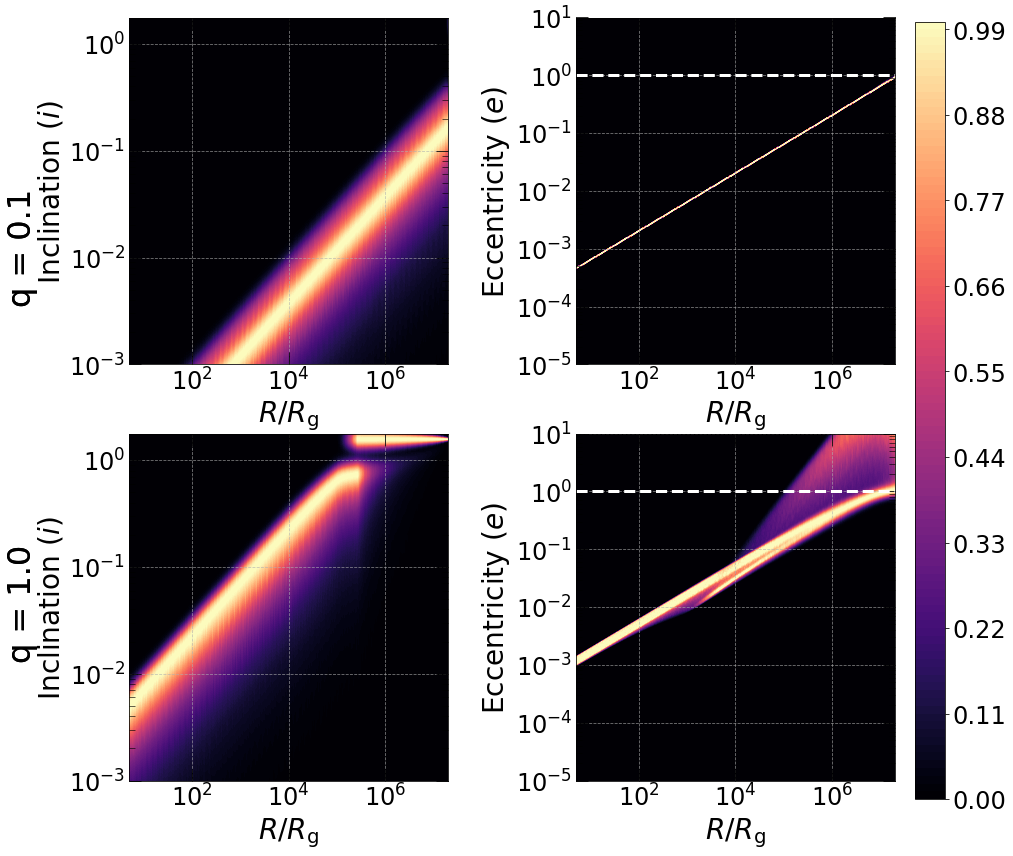}{0.49\textwidth}{(b) $\chi_{1}=0.7,\chi_{2}=0.1$}}
    \caption{Normalized distributions of inclination and eccentricity (from left to right in each panel) for BBH mergers at each radius for different conditions. The left panel (a) shows results for both black holes with dimensionless spin magnitude $\chi=0.7$. The right panel shows results for one black hole (the heavier one) with dimensionless spin $\chi =0.7$, and the second black hole with weaker spin $\chi=0.1$. In both panels the top (bottom) row shows a BBH with mass ratio $q=0.1$ ($q = 1$). The white dashed line in the eccentricity panels corresponds to $e=1$; all simulations above the line escape the SMBH system due to the kick. In the inclination panels the escaped cases are shown to have inclination of exactly $\pi/2$ (note there is a gap between that and the actually highest inclination).}
    \label{fig : stats}
\end{figure}

More specifically, in the main text, we assume that both spin-orbit misalignment angles $\iota_{1,2}$ can be uniformly drawn from $0 \le \cos(\iota_{1,2}) \le \cos(30^\circ)$.  We assume that the orbital angular momentum vector is parallel to the AGN angular momentum vector.  We use spin magnitudes $\chi_1 = \chi_2 = 0.7$, as is appropriate for comparable-mass merger products, and $q=1$ (a conservative assumption for our purposes, since this comes close to maximizing the kick velocity for systems with spinning components).  However, we also explore variations in our assumptions regarding $q$ and $\chi_1, \chi_2$ in Fig. \ref{fig : stats}.

At each galactocentric orbital radius in a given AGN disk model, we compute $10^4$ recoil velocities, which are then added to the orbital Keplerian velocity $v_{\rm K}=\sqrt{GM/r} = c \sqrt{R_{\rm g}/r}$ of the binary:
\begin{equation}
    \mathbf{v} = v_{\rm K} \hat{y} + \mathbf{v}_{\rm recoil}
\end{equation}
From here, we calculate the new inclination, eccentricity, and pericenter. We can calculate the relevant orbital parameters independently of the SMBH mass:
\begin{align}
    e &= \left| \left( \frac{|\mathbf{v}|^2}{GM}-\frac{1}{|\mathbf{r}|}\right)\mathbf{r} - \frac{\mathbf{v}\cdot\mathbf{r}}{GM}\mathbf{v} \right| = \left| \left( \frac{|\mathbf{v}|^2}{c^2} \left( \frac{r}{R_{\rm g}} \right) - 1 \right) \hat{r} - \frac{r}{R_{\rm g}} \frac{\mathbf{v}\cdot \hat{r}}{c^2}\mathbf{v} \right| \\
    i &= \arccos \left( \frac{(\mathbf{r} \times \mathbf{v})_z}{|(\mathbf{r} \times \mathbf{v})|} \right) = \arccos \left( \frac{(\hat{x} \times \mathbf{v})_z}{|(\hat{x} \times \mathbf{v})|} \right) \\ 
    \epsilon &= \frac{1}{2}{|\mathbf{v}|^2} - \frac{GM}{r} = \frac{1}{2} |\mathbf{v}|^2 - c^2 \left( \frac{r}{R_{\rm g}} \right)^{-1} \\
    a_{\rm g} &= \frac{a}{R_{\rm g}}  = - \frac{GM}{2\epsilon}/R_{\rm g} = - \left( \frac{1}{2} \frac{|\mathbf{v}|^2}{c^2} - \left( \frac{r}{R_{\rm g}} \right)^{-1} \right)^{-1} 
\end{align}

Here $e$, $i$, $\epsilon$, and $a_{\rm g}$,  are the eccentricity, inclination,  specific energy and  semi-major axis, respectively (the latter is in $R_{\rm g}$ units). 
For the scenario that produces the strongest gravitational kicks ($q=0.1,\chi_{1,2}=0.7$), we analyzed how many objects get ejected from the supermassive black hole (SMBH) system at different distances
 We found that all mergers within  $R/R_{\rm g} \lesssim 1.8 \times 10^4$ will remain bound to the system. Beyond this distance, the ejection rate increases : at about $\sim 4.4 \times 10^4 R_{\rm g}$, 10\% of objects are ejected; at $\sim 10^5  R_{\rm g}$, 30\% are ejected; and at $\sim 5\times  10^5  R_{\rm g}$, 65\% of objects are ejected from the system.

\subsection{Timescales}
\label{app : timescales}

In this section we enumerate and evaluate different timescales that are relevant for BHs merging in AGN migration traps.

\subsubsection{AGN lifetime} 
The AGN lifetime is calculated from duty cycle arguments using empirical AGN demographic data presented in \cite{Aird+2018}.  Specifically, we use the duty cycle probability (as a function of Eddington ratio $\epsilon_{\rm Edd}$, redshift $z$ and mass $M$), $d\delta_{\rm duty}/d \log_{10} \epsilon_{\rm Edd}$.  To obtain an upper limit on an AGN lifetime we use the "effective" lifetime calculation in \cite{HopkinsHernquist2009}: 
\begin{equation}
    t_{\rm AGN}(>\epsilon_{\rm Edd}) = t_{\rm H}(z)  \delta_{\rm duty} (>\epsilon_{\rm Edd}|z,M) \ ,
\end{equation}
where $t_{\rm H}$ is the Hubble time at redshift $z$.  
We note that the data in \cite{Aird+2018} is given as a function of galaxy stellar mass, which we relate to the SMBH mass first using a bulge-to-total mass scaling relationship \citep{vanVelzen2018} and then using a bulge mass to SMBH mass relationship \citep{KormendyHo2013}. 

We emphasize that because $t_{\rm AGN}$ is obtained from duty cycle arguments, it really reflects the total time during the current age of the Universe that a galaxy has been in AGN mode, but it does not differentiate between a single AGN episode of duration $t_{\rm AGN}$ or ten discrete episodes of duration $t_{\rm AGN}/10$; it is therefore an upper limit on the episodic lifetime.

\subsubsection{Realignment times}
BHs kicked onto inclined orbits (e.g., by GW recoil) will gradually realign into the AGN disk.  Given a specific AGN profile, we calculate how long the realignment process takes using the estimates of \cite{GenerozovPerets2023}. Specifically, we assume that realignment occurs at a time:
\begin{equation}
t_{\rm align} = \min \left[ t_{\rm align}^{\rm GDF},  t_{E}^{\rm GDF} \right]
\end{equation}
where $t_{\rm align}^{\rm GDF}$ is the characteristic alignment time for gas dynamical friction calculated by the change to the $z$ component of momentum.   For high inclination the rate of change of energy is more accurate, so the dynamical friction alignment timescale is $t_{E}^{\rm GDF}$.
Note that \cite{GenerozovPerets2023} also include ram pressure cases, where the relative velocity is much higher than the escape velocity of the body. Because in this work we care more about black holes (where the escape velocity is the speed of light) we always care only about dynamical friction. The specific expressions for each of these timescales are \citep{GenerozovPerets2023}:
\begin{align}
t_{\rm align}^{\rm GDF} &= k_o \frac{v_{\mathrm{c}}(a)^3}{4 \pi G^2 \rho_g(a) m_* \ln \Lambda}\left(\frac{v_{\rm rel }}{v_{\mathrm{c}}}\right)^3\left(\frac{P_{\rm orb }}{t_{\rm cross }}\right)\left(\frac{r}{a}\right)^\gamma \\
t_{E}^{\rm GDF} &= k_2 \frac{v_{\rm c}(a)^3}{4 \pi G^2 \rho_{\rm g}(a) m_{\star} \ln(\Lambda)} \left( \frac{v_{\rm rel}}{v_{\rm c}} \right) \left( \frac{P_{\rm orb}}{t_{\rm cross}} \right) \left( \frac{r}{a} \right)^{\gamma} \ .
\end{align}
Here $ a,r,i,v_{\rm c}(a),P_{\rm orb},m_\star $ are the semimajor-axis, pericentr (that includes argument of periapsis),inclination, circular orbit velocity, orbital period, and mass, respectively, of the body whose realignment timescale we are calculating. $\rho_{\rm g}, h, \gamma$ are the local density, scale-height and adiabatic index of the disk. $t_{\rm cross}\equiv \frac{2(h / r) r}{v_z}$ is the vertical crossing time. $\ln \Lambda$ and $v_{\rm rel}$  are the Coulomb logarithm and the relative velocity between the body and the gas, respectively.   Fitted choices of  the prefactors are : $k_0=0.18,k_2 =1$.

\subsubsection{Binary-single scattering and merger timescale}
\label{app : 2+1 merger}
We estimate the timescale for merger if the evolution of the binary in the migration trap is driven exclusively by dynamical interaction with a third BH also orbiting in the trap and by gravitational-wave emission.  
 
We consider a BH-BH binary with component masses $m_{\rm a}$ and $m_{\rm b}$ (total mass $m_{\rm B}=m_{\rm a}+m_{\rm b}$) that orbits around an SMBH with mass $M$ at distance $R$.  Also orbiting around the SMBH is a third BH of mass $m_{\rm s}$, at a distance $R+\delta$ from the SMBH.  All SMBH-centric orbits are assumed to be circular, both for simplicity and due to the effects of gas damping \citep{FairbairnRafikov24}.  Any time the third mass is azimuthally aligned with the binary, we refer to the situation as a ``flyby.'' The period between flybys is
\begin{equation}
    \Delta P_{\rm fly}  = \frac{2 \pi (R+\delta) }{\Delta v} \approx  \frac{4 \pi R}{v_{\rm K}} \left( \frac{\delta}{R} \right)^{-1}.
\end{equation}
The relative velocity between the binary and the tertiary is $\Delta v \approx (1/2) v_{\rm K}(R) \left( \delta / R \right)$. We are interested in understanding the rate of strong binary-single scatterings, in which the tertiary passes within a radius $a$ (i.e. the internal semimajor axis) of the binary.  For a large enough $a$, any flyby within the migration trap will result in a strong interaction, but if $a$ is small, the probability for a strong scattering is reduced. The critical impact parameter for a strong encounter in the circular and coplanar limit is
\begin{equation}
    \delta_{\rm c} = a \sqrt{1+ \frac{2G m_{\rm tot}}{a \Delta v^2}}. \label{eq:deltaGeneral}
\end{equation}
Here $m_{\rm tot}=m_{\rm a}+m_{\rm b}+m_{\rm s}$. Since $\Delta v^2$ is itself a function of $\delta_{\rm c}$, Eq. \ref{eq:deltaGeneral} is a quadratic that can be solved as
\begin{equation}
    \delta_{\rm c}^2 = \frac{1}{2} \tilde{a}^2 R^2 \left( \frac{m_{\rm B}}{M} \right)^{2/3} \left(1 + \sqrt{1+32 \tilde{a}^{-3} \frac{m_{\rm tot}}{m_{\rm B}}} \right).
\end{equation}
Here we have normalized the binary semimajor axis as a fraction of the Hill radius $R_{\rm H} = R (m_{\rm B}/M)^{1/3}$ so that $\tilde{a} = a / R_{\rm H} \le 1$.  In all physical cases of interest, this solution can be expressed approximately as the gravitationally focused limit,
\begin{equation}
    \delta_{\rm c}^{\rm f} = 2^{3/4} \tilde{a}^{1/4} \left( \frac{m_{\rm tot}}{M} \right)^{1/4} \left( \frac{m_{\rm B}}{M} \right)^{1/12} R,
\end{equation}
where the superscript ``f'' denotes that we are using the gravitationally focused approximation for $\delta_{\rm c}$.  Hence the relative velocity for the first flybys that produce strong encounters is
\begin{equation}
    \Delta v_{\rm c} = 2^{-1/4} \tilde{a}^{1/4} v_{\rm K}(R) \left( \frac{m_{\rm tot}}{M} \right)^{1/4} \left( \frac{m_{\rm B}}{M} \right)^{1/12} .
\end{equation}

We now wish to estimate the timescale between strong encounters, $T_{\rm s}$.  
We assume that migration brings the companion to $\delta \sim \delta_{\rm c}$ on the migration timescale, which is the rate-limiting step, but does not subsequently affect the system (a conservative assumption, since otherwise a strong 2+1 interaction would happen even sooner).  We thus approximate  $\Delta v = \Delta v_{\rm c}$.  If the tertiary is on a sufficiently coplanar orbit, then $T_{\rm s} = \Delta P_{\rm fly}$ and is given deterministically by 
\begin{equation}
T_{\rm s, d}^{\rm f} = \frac{2 \pi R}{v_{\rm K}} \left(\frac{2}{\tilde{a}} \right)^{1/4} \left(\frac{m_{\rm tot}}{M} \right)^{-1/4}  \left(\frac{m_{\rm B}}{M} \right)^{-1/12}.
\end{equation}
However, if the tertiary is on a sufficiently inclined orbit (with inclination angle $i$), then not every flyby will produce an impact parameter $< \delta_{\rm c}$; this will only be achieved in a fraction of flybys $\approx \delta / R \sin(i)$.  In this limit, the time between strong scatterings is stochastic, but the average time is
\begin{equation}
T_{\rm s, s}^{\rm f} =  \frac{2 \pi R}{v_{\rm K}} \left(\frac{1}{2\tilde{a}} \right)^{1/2} \sin(i) \left(\frac{m_{\rm tot}}{M} \right)^{-1/2}  \left(\frac{m_{\rm B}}{M} \right)^{-1/6}.
\end{equation}
Eventually, GW emission will dominate subsequent evolution and a merger will happen within a Peters time \citep{Peters1964}:
\begin{equation}
    T_{\rm GW}=\frac{5}{256}\frac{c^{5} a^4}{G^{3} m_{\rm a} m_{\rm b} m_{\rm B} }
    \label{eq: Peters time}.
\end{equation}
We can find the critical binary semimajor axis $a_{\rm GW}$ for which the transition happens if we equate the times $T_{\rm s} =T_{\rm GW}$.  In the deterministic regime, 
\begin{equation}
    \tilde{a}_{\rm GW, d}^{17/4} = 2^{1/4} \frac{512\pi}{5} \left( \frac{R}{R_{\rm g}}\right)^{-5/2} \left( \frac{m_{\rm a} m_{\rm b}}{M^2}\right)  \left(\frac{m_{\rm tot}}{M} \right)^{-1/4}  \left(\frac{m_{\rm B}}{M} \right)^{-5/12}.
        \label{eq : a_GW no probability}
\end{equation}
In the stochastic regime, 
\begin{equation}
    \tilde{a}_{\rm GW, s}^{9/2} = 2^{-1/2} \frac{512\pi}{5} \sin(i) \left( \frac{R}{R_{\rm g}}\right)^{-5/2} \left( \frac{m_{\rm a} m_{\rm b}}{M^2}\right)  \left(\frac{m_{\rm tot}}{M} \right)^{-1/2}  \left(\frac{m_{\rm B}}{M} \right)^{-1/2}.
    \label{eq : a_GW with sin}
\end{equation}
In most situations, a binary will begin its hardening in the deterministic regime and eventually transition to the stochastic one if $\sin(i) \sim H/R$ (it may remain deterministic if $\sin(i) \ll H/R$).  In either regime, the strong scattering and GW inspiral times are both very short.  Using Eqs. \ref{eq : a_GW no probability} and \ref{eq : a_GW with sin} together with   \cref{eq: Peters time} we get:
\begin{align}
    \frac{T_{\rm GW,d}}{\rm year}&=2.43\times10^{5}\left(\frac{M}{10^{8}M_{\odot}}\right)\left(\frac{R}{10^{4}R_{{\rm g}}}\right)^{28/17}\left(\frac{m_{{\rm a}}m_{{\rm b}}m_{B}}{10^{-21}M^{3}}\right)^{-1/17}\left(\frac{m_{{\rm tot}}}{10^{-7}M}\right)^{-4/17}, \\ 
    \frac{T_{\rm GW,s}}{\rm year}&=10^{7}\sin^{8/9}(i)\left(\frac{M}{10^{8}M_{\odot}}\right)\left(\frac{R}{10^{4}R_{{\rm g}}}\right)^{16/9}\left(\frac{m_{{\rm a}}m_{{\rm b}}m_{B}}{10^{-21}M^{3}}\right)^{-1/9}\left(\frac{m_{{\rm tot}}}{10^{-7}M}\right)^{-4/9},
\end{align}
with the total merger timescale only a factor of a few longer, Eq.~(\ref{eq: 2+1 merger time}).
This indicates that the ``rate-limiting step'' for scattering-driven mergers will be the arrival of tertiaries into the migration trap (see Figure \ref{fig: realignment_time}).

\section{Higher Density AGN Disks}
\label{app : additional results}

\begin{figure}[htbp]
    \centering

    \begin{minipage}{0.45\textwidth}

        \centering
        \includegraphics[width= \linewidth]{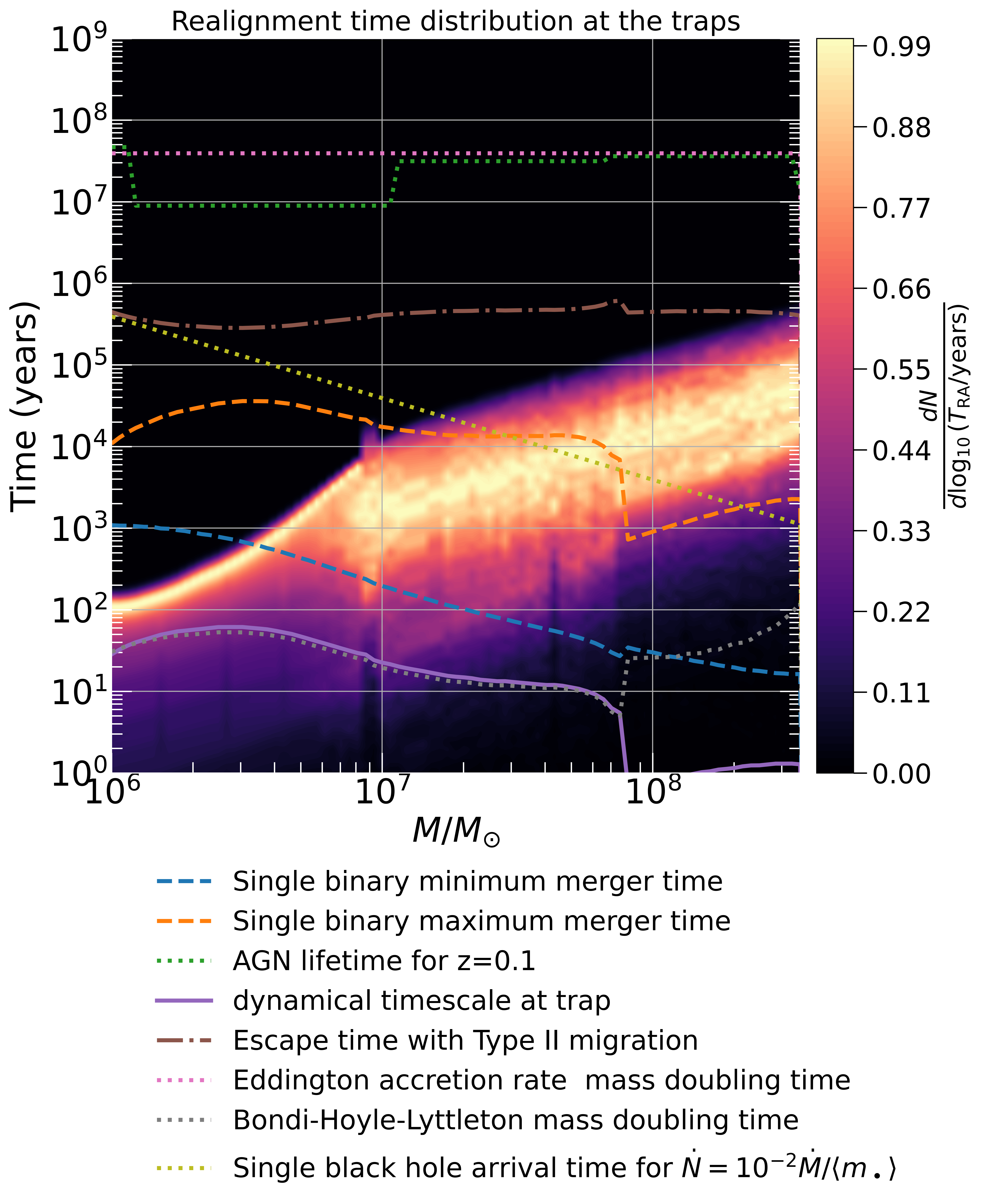}
        \caption{Timescales and re-alignment times for $\alpha=0.01$. Notation is the same as in Fig. \ref{fig: realignment_time} (in which $\alpha=0.1$ was used). In the higher-$\Sigma$ conditions of this low-$\alpha$ disk, we see much shorter re-alignment and Type II escape times.}
        \label{fig: realignment_time, alpha =0.01}
    \end{minipage}
\end{figure}
In our fiducial models, we assume the value $\alpha=0.1$ for the dimensionless Shakura-Sunyaev viscosity parameter, a choice in reasonable agreement with global magnetohydrodynamic simulations of the inner regions of AGN accretion disks \citep{Jiang+19}.  However, the value of $\alpha$ is famously controversial, and some simulations find substantially lower values of $\alpha \sim 0.01$ (e.g. \citealt{Davis+10}).  In this Appendix, we consider the higher gas densities that would occur in AGN disks with $\alpha=0.01$.  Figures  \ref{fig: realignment_time, alpha =0.01},  \ref{fig: mass_for_gap, alpha =0.01}, and \ref{fig: mass_for_escape, alpha =0.01} are re-calculations of Figs. \ref{fig: realignment_time} , \ref{fig: mass_for_gap}, and \ref{fig: mass_for_escape} (respectively), with $\alpha$ reduced from our fiducial value of $0.1$ to a new value of $0.01$.

In Fig. \ref{fig: realignment_time, alpha =0.01}, we see that a reduction in disk viscosity reduces the realignment times of kicked BHs by about an order of magnitude.  As in the $\alpha=0.1$ case, realignment times are typically far longer than the disk dynamical time at the trap location, far shorter than the AGN lifetime or the Salpeter time, and at least somewhat shorter than the maximum merger time due to repeated binary-single scatterings, as well as the ``escape time'' due to Type II migration.  Unlike the $\alpha=0.1$ case, there is now a large portion of parameter space ($M \lesssim 10^7 M_\odot$) where the {\it minimum} merger time from binary-single scatterings is now longer than the realignment time.  In general, the timescale hierarchies for trap dynamics appear robust to the choice of $\alpha$.

Fig. \ref{fig: mass_for_gap, alpha =0.01}, by contrast, shows more significant dependence on $\alpha$.  In particular, the $\alpha=0.01$ case has larger gap-opening masses at the location of the migration trap, with typical $M_{\rm gap} \sim 30 M_\odot$.  This contrasts with the $\alpha=0.1$ case from Fig. \ref{fig: mass_for_gap}, where $M_{\rm gap} \lesssim 10 M_\odot$ for $M < 10^7 M_\odot$.  In other words, higher disk gas densities that  delay gap opening and create more space for hierarchical growth into the pair instability gap. This effect is especially relevant to the "Island" migration traps that are much closer to the center with much higher densities. This effect is seen also  in Fig. \ref{fig: mass_for_escape, alpha =0.01}, we see a sudden jump to extreme masses (lower right side).  In Fig. \ref{fig: mass_for_escape, alpha =0.01}  we also see that somewhat lower AGN luminosities are capable of hosting growth into the pair instability mass gap.  

\begin{figure}[htbp]
    \hfill
    \begin{minipage}{0.45\textwidth}
        \centering
        \includegraphics[width=\linewidth]{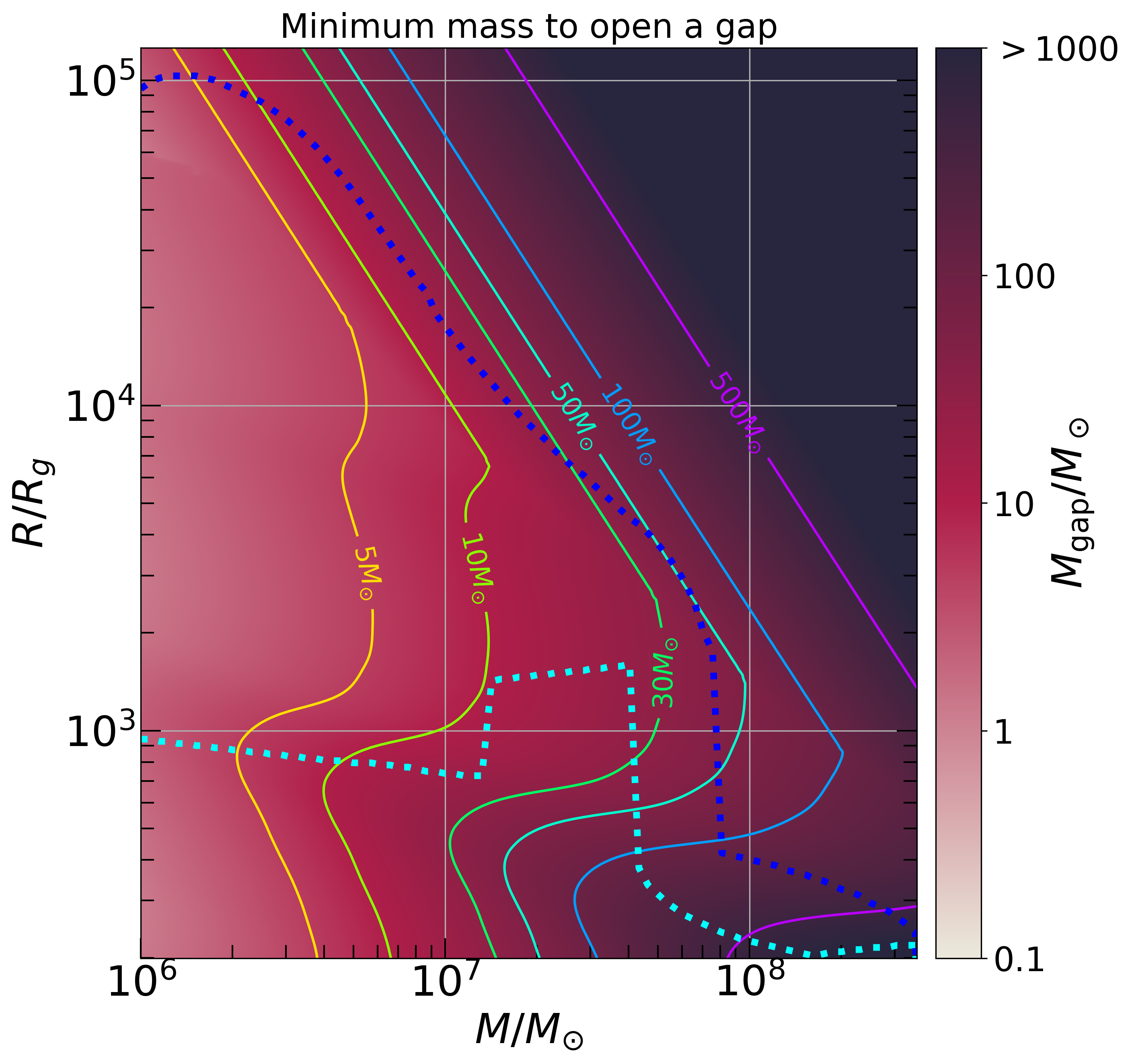}
        \caption{The color-coded minimum mass for opening a gap, $M_{\rm gap}$, computed for an ensemble of $\alpha=0.01$ AGN disks with varying central mass $M$. Notation is the same as in Fig. \ref{fig: mass_for_gap} (where $\alpha=0.1$ was used). One notable change from the high-$\alpha$ case is that here, we see regions ($M\sim 5\times 10^7 M_\odot$) where ``gap-closing'' traps exist, i.e. where $M_{\rm gap}$ increases at radii beneath the linear trap.}
        \label{fig: mass_for_gap, alpha =0.01}
    \end{minipage}
        \begin{minipage}{0.45\textwidth}
    \centering
    \includegraphics[width=\textwidth]{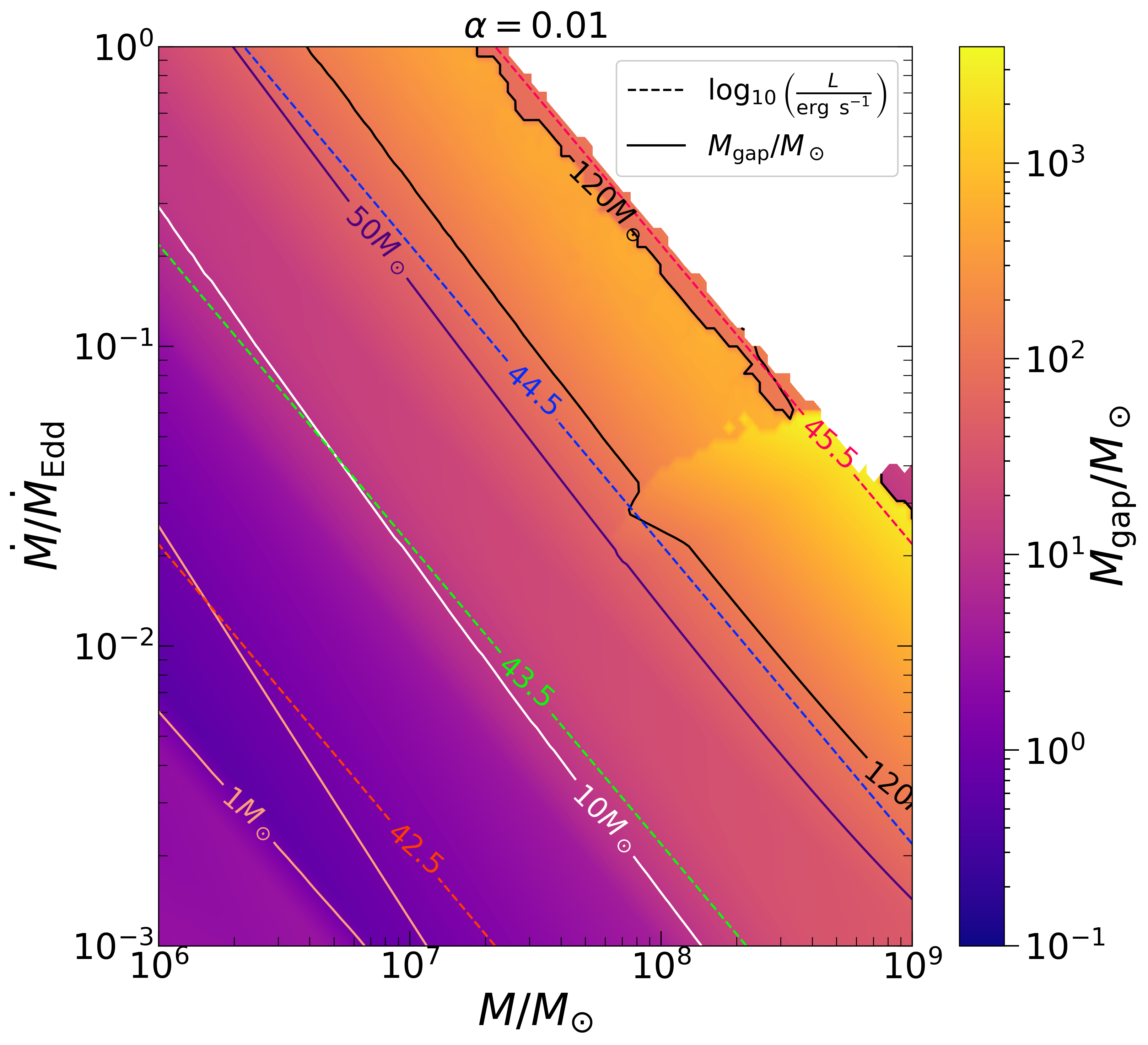}
    \caption{The minimum mass for escaping the migration trap via gap opening (i.e. the maximum gap opening mass $M_{\rm gap}$ between the anti-trap and trap) for $\alpha=0.01$. Notation is the same as in Fig. \ref{fig: mass_for_escape} (where $\alpha=0.1$ was used). The higher-$\Sigma$ disks shown here feature, in some parts of parameter space, much larger maximum $M_{\rm gap}$ values, peaking at $\sim 4\times 10^3 M_{\odot}$ for high $M$ and low Eddington ratio $\dot{M}/\dot{M}_{\rm Edd}$.}
    \label{fig: mass_for_escape, alpha =0.01}
    \end{minipage}
\end{figure}

\end{document}